\documentstyle[12pt,titlepage]{article}
 \input epsfig.sty

.5 scaled \magstep4
\font\medio=cmr9.5 scaled \magstep2
\outer\def\beginsection#1\par{\medbreak\bigskip
      \message{#1}\leftline{\bf#1}\nobreak\medskip
\vskip-\parskip
      \noindent}

\def\laq{\raise 0.4ex\hbox{$<$}\kern -0.8em\lower 0.62
ex\hbox{$\sim$}}
\def\gaq{\raise 0.4ex\hbox{$>$}\kern -0.7em\lower 0.62
ex\hbox{$\sim$}}

\begin{document}
\bibliographystyle {unsrt}

\titlepage

\begin{flushright}
CERN-PH-TH/2008-070
\end{flushright}

\vspace{15mm}
\begin{center}
{\Large CMB polarization induced by stochastic magnetic fields}\\
\vspace{15mm}
 
Massimo Giovannini$^{a,c}$ and Kerstin E. Kunze$^{b,c}$\\

\vskip2.cm

{\sl $^a$INFN, Section of Milan-Bicocca, 20126 Milan, Italy}
\vskip 0.2cm 
{\sl $^b$ Departamento de F\'\i sica Fundamental, \\
 Universidad de Salamanca, Plaza de la Merced s/n, E-37008 Salamanca, Spain}
\vskip 0.2cm 
{\sl $^c$  Department of Physics, Theory Division, CERN, 1211 Geneva 23, Switzerland}
\vspace{6mm}

\end{center}

\vskip 2cm
\centerline{\medio  Abstract}
The complete calculation  of the CMB polarization 
observables (i.e. E- and B-modes) is reported within the conventional $\Lambda$CDM paradigm supplemented 
by a stochastic magnetic field. Intriguing perspectives for present and forthcoming CMB polarization 
experiments are outlined.
\noindent

\vspace{5mm}
\vfill 
\newpage
Large-scale magnetism recently  became an 
intriguing triple point where cosmology, astronomy and high-energy astrophysics 
meet for complementary purposes \cite{rev1}. Still we have no clues on its origin. 
 The gravitational instability together with the subsequent 
galactic rotation could amplify a magnetic field spanning a collapsing region 
of the order of the Mpc in comoving units. The magnetic field regularized over such a scale $L$, i.e. 
$B_{\mathrm{L}}$, is not empirically observable  at the epoch of the gravitational collapse 
of the protogalaxy. But the Universe is a good conductor: 
 the magnetic flux and  helicity  are approximately conserved implying that
 large-scale magnetic fields could have been already present at the time 
when photons last-scattered electrons and ions, i.e., according to the WMAP 5-year 
data \cite{WMAP5}, at a redshift $z_{\mathrm{dec}}\simeq 1090$.

Intriguing effects related to tangled magnetic fields 
have been discussed with semi-analytical methods (see, in particular, \cite{bs1}). More recently 
the impact of large-scale magnetic fields on scalar modes of the geometry have been 
addressed \cite{mg1} and a dedicated numerical approach has 
been devised \cite{gk}. The complete calculation of the polarization angular power spectra 
(i.e., specifically, the EE, TE and BB angular power spectra)
is here reported, for the first time, when the conventional $\Lambda$CDM paradigm is complemented 
by a stochastic magnetic field (i.e., according to the terminology of \cite{gk}, m$\Lambda$CDM scenario). 

In short the main theoretical impasse is the following. The large-scale 
description of temperature anisotropies demands a coarse grained (one-fluid) approach for the 
electron-ion system: this is the so called baryon fluid which is treated (with no exceptions) as a single 
fluid in popular Boltzmann solvers such as COSMICS \cite{mb} and CMBFAST \cite{cmbfast}. On the other hand the dispersive propagation of electromagnetic disturbances demands to treat separately electrons and ions, at least at high frequencies. 
It is appropriate to start from the  Vlasov-Landau equation written in the form:
\begin{equation}
\frac{\partial f_{\pm}}{\partial \tau} + v^{i} \frac{\partial f_{\pm}}{\partial x^{i}} \pm e 
( E^{i} + v_{j} B_{k} \epsilon^{j\,k\,i}) \frac{\partial f_{\pm}}{\partial q^{i}}+ \frac{1}{2} h_{ij}' q^{i} \frac{\partial f_{\pm}}{\partial q^{j}} = 
{\mathcal C}_{\mathrm{coll}}.
\label{VL}
\end{equation}
where $\vec{v} = \vec{q}/\sqrt{m^2 a^2 + q^2}$ is the  comoving three velocity, $\vec{q}$ is the comoving three-momentum and 
$\tau$ is the conformal time arising, in the line element, as $ds^2 = a^2(\tau)\{d\tau^2 - [\delta_{ij} - h_{ij}(\vec{x},\tau)]dx^{i} dx^{j}\}$. The prime denotes a derivation with respect to $\tau$. The rescaled electromagnetic fields are denoted as 
$\vec{E} = a^2 \vec{{\mathcal E}}$ and as $\vec{B} = a^2 \vec{{\mathcal B}}$.
By choosing the plus (minus) sign in  Eq. (\ref{VL}), the evolution equation for the one-body distribution function 
$f_{\pm}(\vec{x}, \vec{q}, \tau)$ of the ions (electrons) can be obtained\footnote{The velocity-configuration space naturally arises since ions and electrons are all non-relativistic. Consequently, the comoving three-momentum is given by 
$\vec{q} = m a \vec{v}$ for each of the two charged species. The quasi-equilibrium distribution for 
electrons and protons is Maxwellian and the strength of Coulomb scattering guarantees 
$T_{\mathrm{e}} \simeq T_{\mathrm{i}} \simeq T$.  For relativistic (neutral) species $q^{i} = n^{i} q$. The equilibrium 
distribution for neutrinos and photons will be, respectively,  Fermi-Dirac and Bose-EInstein.}. 
In the electron-ion system ${\mathcal C}_{\mathrm{coll}}$ is provided by Coulomb scattering.

In the limit $e\to 0$, Eq. (\ref{VL}) describes  the evolution of neutral species. If ${\mathcal C}_{\mathrm{coll}}=0$, 
 Eq. (\ref{VL}) leads, below the MeV, to the well known evolution equation for the reduced phase space 
distribution of the neutrinos in the synchronous gauge\footnote{Recall, for this 
purpose, that the scalar fluctuation of the geometry $h_{ij}(\vec{x},\tau)$ carries two degrees of freedom, i.e., in Fourier space, 
$h_{ij}(\vec{k},\tau) = [\hat{k}_{i} \hat{k}_{j} h(k,\tau) + 2 \xi(k,\tau)( 3 \hat{k}_{i} \hat{k}_{j} - \delta_{ij})]$.} \cite{mg1,gk}:
\begin{equation}
{\mathcal F}_{\nu}' + i k\mu {\mathcal F}_{\nu} = 2 \mu^2 ( h' + 6 \xi') - 4 \xi',\qquad \mu = \hat{k}\cdot\hat{n}.
\label{VL1}
\end{equation}
where, as in the conventional $\Lambda$CDM models the neutrinos are massless and, consequently, 
$v^{i} = q^{i}/|\vec{q}|= n^{i}$. 
The evolution equations of the brightness perturbations of the intensity (i.e. $\Delta_{\mathrm{I}}$) 
and of the polarization (i.e. $\Delta_{\mathrm{Q}}$ 
and $\Delta_{\mathrm{U}}$),  can be derived from Eq. (\ref{VL}) (always in the limit $e\to 0$) by identifying 
${\mathcal C}_{\mathrm{coll}}$ with the electron-photon collision term when the energy of the photons 
is parametrically smaller then the electron mass and when the electron recoil is neglected:
\begin{eqnarray}
&&  \Delta_{\mathrm{I}}' +( i k \mu + \epsilon') \Delta_{\mathrm{I}} = - \biggl[ \xi' - \frac{\mu^2}{2}( h' + 6 \xi')\biggr] +
\epsilon' \biggl[ \Delta_{\mathrm{I}0} + \mu v_{\mathrm{b}} - \frac{3 \mu^2 - 1}{4}(\mu) {\mathcal S}\biggr],
\label{VL1a}\\
&&\Delta_{\mathrm{Q}}' + ( i k\mu + \epsilon') \Delta_{\mathrm{Q}} = \frac{3 \epsilon' }{4}(1 - \mu^2) {\mathcal S},\qquad 
\Delta_{\mathrm{U}}' + ( i k\mu + \epsilon') \Delta_{\mathrm{U}} =0, 
\label{VL2}
\end{eqnarray}
where $\epsilon'= a x_{\mathrm{e}}\,\tilde{n}_{\mathrm{e}}\,\sigma_{\mathrm{Th}}$ is the differential optical depth 
defined in terms of the ionization fraction $x_{\mathrm{e}}$, of the electron density, and of the Thompson cross section 
$\sigma_{\mathrm{Th}}$; $v_{\mathrm{b}} = \theta_{\mathrm{b}}/(ik)$ is the baryon velocity to be defined in a moment. The source 
term in Eqs. (\ref{VL1a}) and  (\ref{VL2}), i.e. ${\mathcal S} =( \Delta_{\mathrm{I}2}+ \Delta_{\mathrm{Q}2} + \Delta_{\mathrm{Q}0})$, contains
the quadrupole of the intensity of the radiation field, $\Delta_{\mathrm{I}2}$. In a nutshell, the CMB is linearly polarized (i.e. 
$\Delta_{\mathrm{Q}} \neq 0$) since the amount of linear polarization is proportional, to first-order in the tight-coupling expansion, 
to the quadrupole of the intensity which is, in turn, proportional to the first-order dipole. This reasoning can be generalized to include the effects of the magnetohydrodynamical Lorentz force.
The one-body distributions for electrons and ions enter Maxwell's equations as
\begin{eqnarray}
&&\vec{\nabla}\cdot \vec{E} = 4\pi e \int d^{3}v [f_{+}(\vec{x}, \vec{v}, \tau) - f_{-}(\vec{x}, \vec{v}, \tau)], \qquad \vec{\nabla}\cdot \vec{B} =0,
\label{MX1}\\
&& \vec{\nabla} \times \vec{E}  + \vec{B}' = 0, \qquad \vec{\nabla}\times \vec{B} - \vec{E}'= 
4\pi e \int d^{3}v\,\vec{v}\,[f_{+}(\vec{x}, \vec{v}, \tau) - f_{-}(\vec{x}, \vec{v}, \tau)].
\label{MX2}
\end{eqnarray}
\begin{figure}[tb]
\begin{center}
\begin{tabular}{|c|c|}
      \hline
      \hbox{\epsfxsize = 6.9 cm  \epsffile{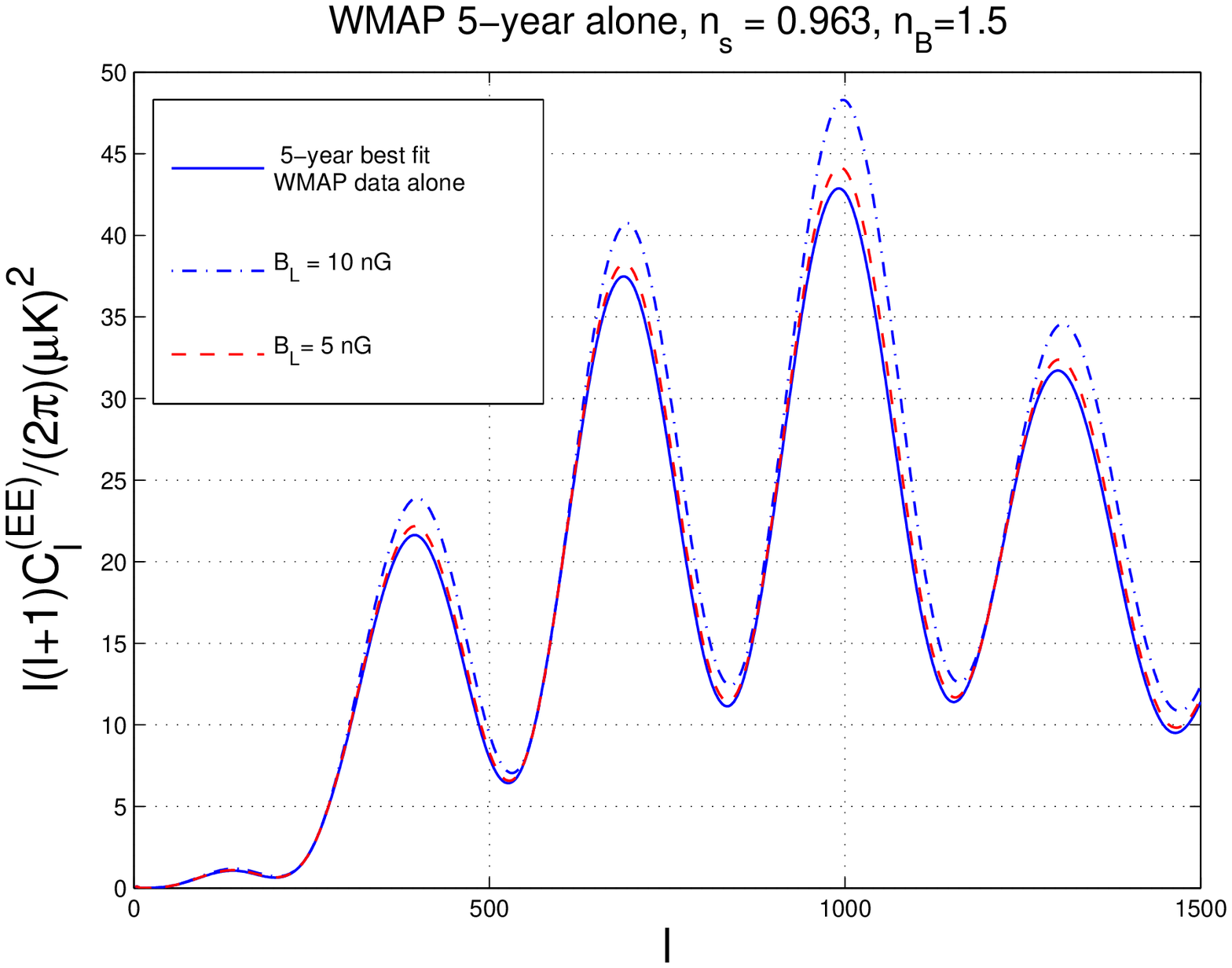}} &
     \hbox{\epsfxsize =  6.9 cm  \epsffile{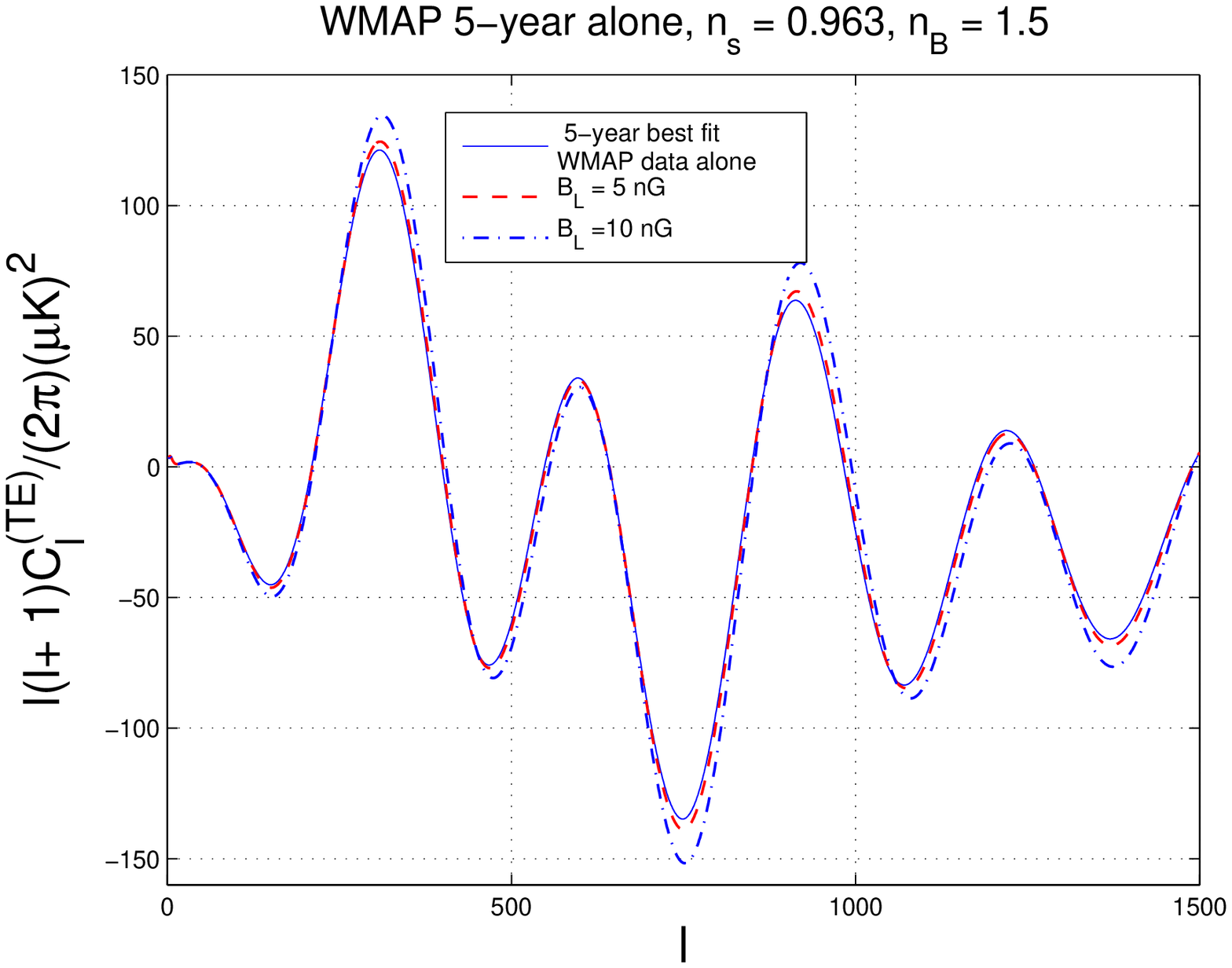}}\\
      \hline
\end{tabular}
\end{center}
\caption[a]{The TE and EE angular power spectra.}
\label{Figure1}
\end{figure}
For length scales much larger than the Debye scale\footnote{The Debye length is defined as $\lambda_{\mathrm{D}} = \sqrt{T/(8 \pi e^2 n_{0})}$ with $n_{0} = a^3\tilde{n}_{\mathrm{e}} \equiv a^3 \tilde{n}_{\mathrm{i}}$ where 
$\tilde{n}_{\mathrm{e}}$ and $\tilde{n}_{\mathrm{i}}$ are the electrons and the ions concentrations.},
 and for  comoving frequencies smaller than the plasma frequency (i.e. 
$\overline{\omega} \ll \overline{\omega}_{\mathrm{pe}}$), Eqs. (\ref{VL1}), (\ref{MX1}) and (\ref{MX2})  reduce 
to an effective one-fluid description where the relevant dynamical variables are given by the centre of mass velocity of  the 
electron-ion system (i.e. $\vec{v}_{\mathrm{b}} = (m_{\mathrm{e}} \vec{v}_{\mathrm{e}} + m_{\mathrm{i}} \vec{v}_{\mathrm{i}})/(m_{\mathrm{e}} + m_{\mathrm{i}})$) and  by 
the total current $\vec{J}$. Ions (i.e. protons) are much heavier than electrons: 
the obtained equations can be expanded 
in powers of $m_{\mathrm{e}}/m_{\mathrm{p}} \ll 1$. These approximations lead to (resistive) magnetohydrodynamics (MHD)
where the electromagnetic disturbances cannot propagate: the total current is solenoidal  (i.e., according to Eq. (\ref{MX2}), 
$4 \pi \vec{J} = \vec{\nabla}\times \vec{B}$) and the electric fields vanish in the baryon rest 
frame with an accuracy determined by the inverse of the Coulomb conductivity.
 The Lorentz force affects anyway the dynamics of the photon-baryon fluid:
\begin{eqnarray}
&&\theta_{\gamma \mathrm{b}}' + \frac{{\mathcal H} R_{\mathrm{b}}}{R_{\mathrm{b}} + 1} \theta_{\gamma\mathrm{b}} = \frac{3}{4 a^4 \rho_{\gamma}} \vec{\nabla}\cdot 
[\vec{J} \times \vec{B}] - \frac{\nabla^2 \delta_{\gamma}}{4 (R_{\mathrm{b}} + 1)}, 
\label{TC1}\\
&& \delta_{\gamma}' = \frac{2}{3} h' - \frac{4}{3} \theta_{\gamma\mathrm{b}}, \qquad \delta_{\mathrm{b}}' =  \frac{h'}{2} - \theta_{\gamma\mathrm{b}},\qquad R_{\mathrm{b}} = \frac{3}{4} \frac{\rho_{\mathrm{b}}}{\rho_{\gamma}},
\label{TC2}
\end{eqnarray}
where, by definition, $\theta_{X} = \vec{\nabla}\cdot \vec{v}_{X}$ is the three divergence of the velocity field and 
$\theta_{\gamma\mathrm{b}} = \theta_{\gamma} = \theta_{\mathrm{b}}$. In Eq. (\ref{TC2}) $\delta_{\gamma}$ 
and $\delta_{\mathrm{b}}$ are the density contrasts of photons and baryons. Both the metric fluctuations and the 
density contrasts of the various species will enter the corresponding Einstein equations whose explicit 
form can be found in \cite{mg1,gk}. 

The comoving (angular) frequency corresponding to the maximum of the CMB 
spectrum  is $\overline{\omega}_{\mathrm{max}} = 2 \pi \nu_{\mathrm{max}}$ where 
$\nu_{\mathrm{max}} = 222.617$ GHz. The comoving plasma frequency is
instead $ \overline{\omega}_{\mathrm{pe}} = 0.285 \,\,\mathrm{MHz}$ for 
$h_{0}^2 \Omega_{\mathrm{b}0}= 0.02273$ (as implied by the best fit to the WMAP 5-year 
data alone).  For CMB photon frequencies  $\overline{\omega} > \overline{\omega}_{\mathrm{pe}}$, Eq. (\ref{VL}) cannot be reduced 
to a one-fluid description. The stochastic magnetic field 
(obeying the one-fluid MHD equations) can then be treated as a background field.
 The propagation of the electromagnetic disturbances is calculated by taking into account the dynamics of ions and electrons 
 within the cold plasma\footnote{The plasma 
parameter \cite{plbook} is  $g_{\mathrm{plasma}}= (V_{\mathrm{D}} n_{0})^{-1}$, i.e. the inverse of the 
number of charge carriers inside the Debye sphere. Around decoupling $g_{\mathrm{plasma}} \simeq 2.3\times  10^{-7} 
\sqrt{x_{\mathrm{e}}}$ for the typical value of the baryonic density implied by WMAP 5-year data.} approximation 
which is rather safe since 
$T_{\mathrm{i}} \ll m_{\mathrm{i}}$ and $T_{\mathrm{e}} \ll m_{\mathrm{e}}$ and $g_{\mathrm{plasma}}\ll 1$.
The initial conditions of the Einstein-Boltzmann hierarchy are given in terms of the so-called magnetized 
adiabatic mode \cite{mg1,gk} which is a solution of the system formed by Eqs. (\ref{TC1})--(\ref{TC2}) 
(as well as by the other MHD and Einstein equations) in the tight-Thompson coupling approximation.
The two sources of inhomogeneity of the system are represented by the stochastic magnetic field (which follows the effective set 
of one-fluid equations) and by the curvature perturbations. The Fourier amplitudes of the  large-scale magnetic field 
will satisfy:
\begin{equation}
\langle B_{i}(\vec{k}) B_{j}(\vec{p}) \rangle = \frac{2\pi^2}{k^3} {\mathcal P}_{\mathrm{B}}(k)P_{ij}(k) \delta^{(3)}(\vec{k} + \vec{p}),\qquad 
{\mathcal P}_{\mathrm{B}} = A_{\mathrm{B}} \biggl(\frac{k}{k_{\mathrm{L}}}\biggr)^{n_{\mathrm{B}} -1}.
\label{BPS}
\end{equation}
where $A_{\mathrm{B}}$ is the spectral amplitude, $n_{\mathrm{B}}$ is the spectral index and $k_{\mathrm{L}}$ is 
the magnetic pivot scale; $P_{ij} = 
(\delta_{ij} - \hat{k}_{i}\hat{k}_{j})$ the transverse projector.
The curvature perturbations will be assigned consistently with the notations 
of Eq. (\ref{BPS}) and, in particular, their power spectrum will be given by 
${\mathcal P}_{{\mathcal R}}(k) = {\mathcal A}_{{\mathcal R}} (k/k_{\mathrm{p}})^{n_{\mathrm{s}} -1}$
where ${\mathcal A}_{{\mathcal R}}$ is the spectral amplitude at the pivot scale $k_{\mathrm{p}} = 0.002\, \mathrm{Mpc}^{-1}$; 
$n_{\mathrm{s}}$ is the spectral index\footnote{For the $\Lambda$CDM paradigm the 5-year WMAP data (alone) imply $n_{\mathrm{s} }= 0.963^{+0.014}_{-0.015}$.}. 
Our numerical code extends the code described in \cite{gk} and it is based on CMBFAST 
\cite{cmbfast}.  As decoupling 
approaches $\theta_{\gamma} \neq \theta_{\mathrm{b}}$ and the linear polarization 
is generated. For frequencies of the CMB photons much larger than $\overline{\omega}_{\mathrm{pe}}$,
dispersive effect come into play (i.e. $\theta_{\mathrm{e}} \neq \theta_{\mathrm{i}}$) and the linear polarization 
is rotated\footnote{There is a second class of dispersive effects
 which is related to the so-called ordinary and extraordinary waves \cite{plbook}.
The latter dispersion relations are insignificant  since the scales of the problem imply that the refractive indices 
are $1$ both for the ordinary and extraordinary waves \cite{gk2}.} 
 leading, ultimately, to the BB angular power spectrum.  Such a rotation 
is proportional to $\hat{n}\cdot\vec{B}$ where $\hat{n}$, as before,
is the direction of the photon momentum. The Larmor radius of the electrons and of the 
ions is much larger than the inhomogeneity scale of the magnetic field: the dynamics of electrons and ions (as well as the dispersion relations \cite{gk2}) can be studied under the guiding centre approximation pioneered by Alfv\'en \cite{ALF1}.  
The  two helicities composing the (linear) CMB polarization 
 propagate with different phase (as well as group) velocities. The Faraday rotation rate 
depends upon the difference $\overline{\omega}[n_{+}(\overline{\omega}) - n_{-}(\overline{\omega})]/2$ where $n_{\pm}(\overline{\omega})$ are the refractive indices of the two circularly polarized waves (one with positive helicity and the other with negative helicity).
\begin{figure}[tb]
\begin{center}
\begin{tabular}{|c|c|}
      \hline
      \hbox{\epsfxsize = 6.9 cm  \epsffile{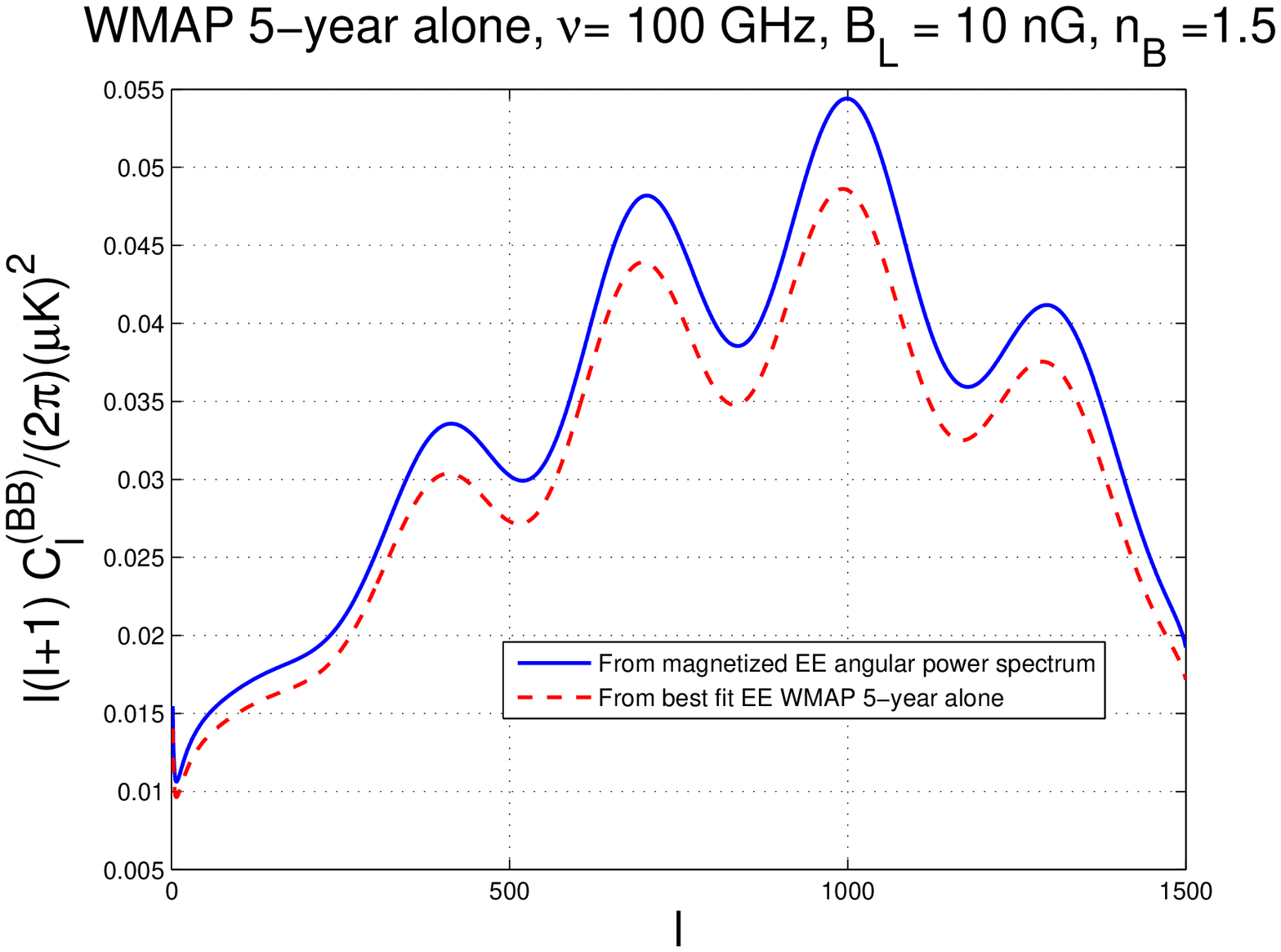}} &
     \hbox{\epsfxsize = 6.9 cm  \epsffile{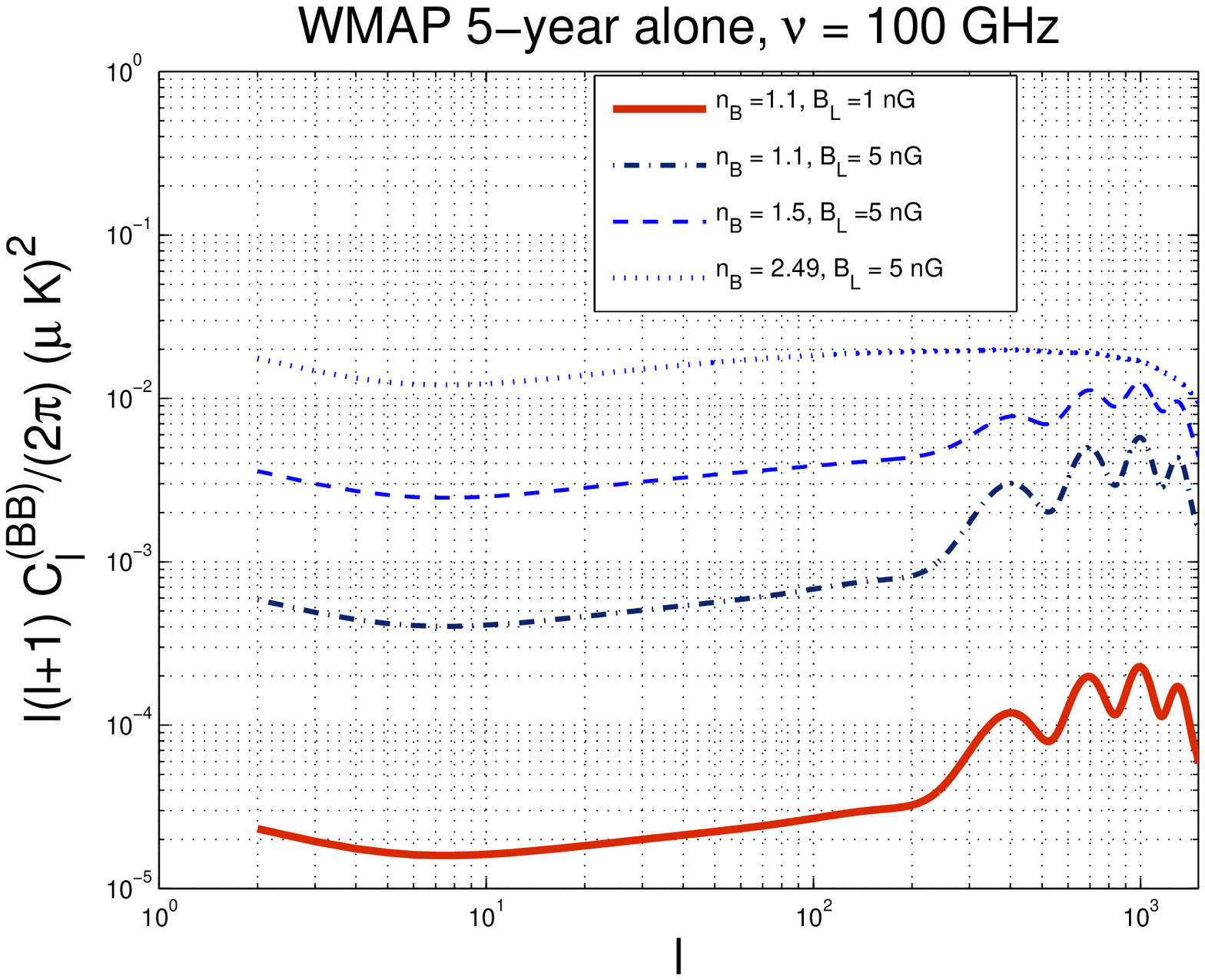}}\\
      \hline
\end{tabular}
\end{center}
\caption[a]{The BB angular power spectra induced by the dispersive propagation 
of the electromagnetic signal in a magnetized background.}
\label{Figure2}
\end{figure}
The heat transfer equations are then supplemented, in this regime, by the 
Faraday rotation rate:
\begin{equation}
\Delta_{\mathrm{Q}}' + n^{i} \partial_{i} \Delta_{\mathrm{Q}} = 2 \epsilon' F(\hat{n}) \Delta_{\mathrm{U}}, \qquad 
\Delta_{\mathrm{U}}' + n^{i} \partial_{i} \Delta_{\mathrm{U}} =-  2 \epsilon' F(\hat{n}) \Delta_{\mathrm{Q}}, 
\label{FR}
\end{equation}
where $F(\hat{n}) = 3/( 16\,\pi^2 \, e) \hat{n} \cdot \vec{B}/\nu^2$ and $\nu$ denotes here the  comoving
frequency. Since the Faraday rate depends upon a stochastic field\footnote{It is here assumed that 
spatial isotropy is unbroken (as observations seem to indicate). A {\em uniform} magnetic field (such as the one assumed 
in \cite{F1}) would break spatial isotropy. If the magnetic field breaks spatial isotropy the TB (and possibly EB) power 
spectra will be present. In the present case the latter power spectra vanish.} it will also be characterized 
by a power spectrum whose explicit form  depends upon the spectral amplitude and slope 
of the magnetic field \cite{F2,gk2}. 
In Fig. \ref{Figure1} the EE and TE angular power spectra are reported in the case of the WMAP 5-year data. 
The TE and EE angular are defined from the corresponding expansion coefficients by recalling that, in the present 
notations, ${\mathcal M}_{\pm}(\hat{n}) =
 \Delta_{\mathrm{Q}}(\hat{n}) \pm i \Delta_{\mathrm{U}}(\hat{n}) = \sum_{\ell \, m} a_{\pm2,\,\ell\, m} \, _{\pm 2} Y_{\ell\, m}(\hat{n})$
 where $_{\pm 2} Y_{\ell\, m}(\hat{n})$ are the spin-2 spherical harmonics. In terms of $a_{\pm2,\,\ell\, m}$ 
 the E-mode and the B-mode are given by $a^{(\mathrm{E})}_{\ell\, m} = - (a_{2,\,\ell m} + a_{-2,\,\ell m})/2$ and by 
 $ a^{(\mathrm{B})}_{\ell\, m} =  i (a_{2,\,\ell m} - a_{-2,\,\ell m})/2$.
\begin{figure}[tb]
\begin{center}
\begin{tabular}{|c|c|}
     \hline
     \hbox{\epsfxsize = 6.9 cm  \epsffile{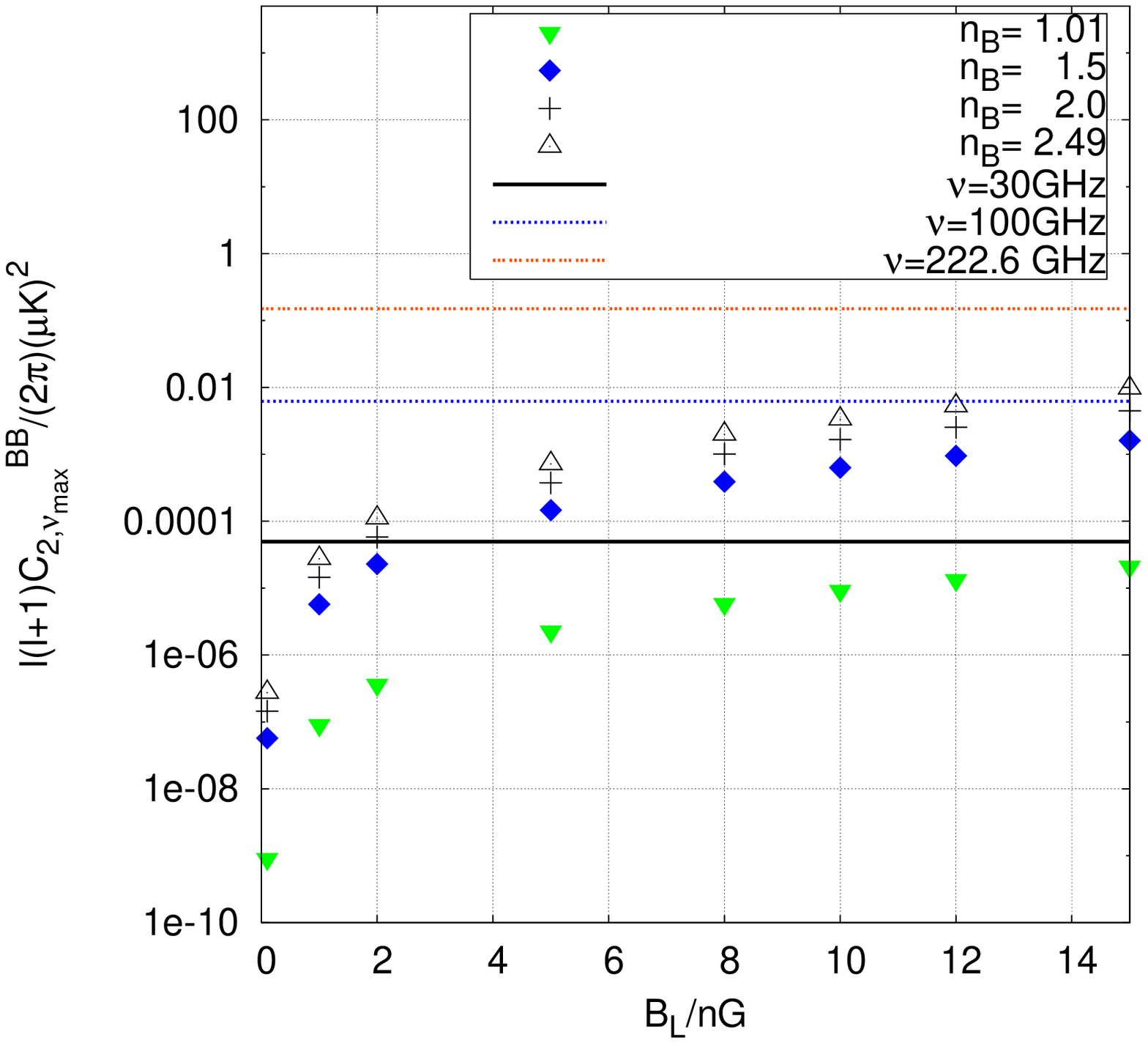}} &
     \hbox{\epsfxsize = 6.9 cm  \epsffile{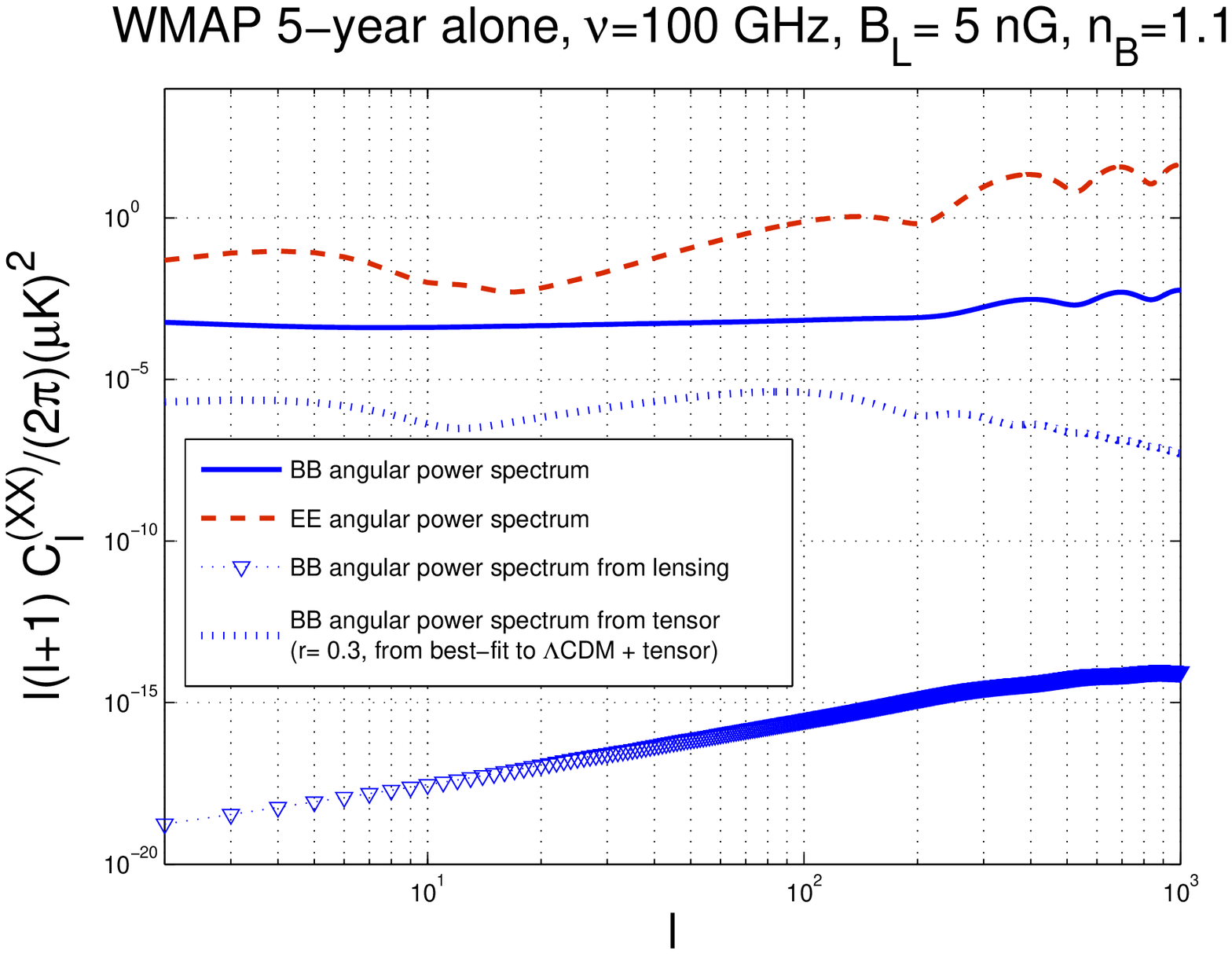}}\\
     \hline
\end{tabular}
\end{center}
\caption[a]{The bound on the BB angular power spectrum from the WMAP 5-year data (plot at the left). At the right, the
magnetized EE and BB power spectra are compared to the B-modes expected from lensing and 
from the tensor mores in the case $r= 0.3$.}
\label{Figure3}
\end{figure}
In Fig. \ref{Figure1} the spectral index as well as all the other parameters of the underlying 
$\Lambda$CDM model have been fixed  to the best-fit value of the WMAP-5year data alone.
In both plots of Fig. \ref{Figure1} the best fit is illustrated with the full lines.
The angular power spectrum of Faraday rotation is defined 
as $C_{\ell}^{(\mathrm{F})} \delta_{\ell \ell'\,m m'} = \langle a_{\ell\, m}^{*} a_{\ell'\,m'}\rangle$
where $a_{\ell\,m} = \int d\Omega_{\hat{n}} Y_{\ell\,m}(\hat{n})^{*} F(\hat{n})$.
In terms of $C_{\ell}^{(\mathrm{F})}$ the autocorrelation of the B-mode will be given by 
\begin{equation}
C_{\ell}^{(\mathrm{BB})} = \sum_{\ell_{1},\ell_{2}} {\mathcal G}(\ell_{1},\ell_{2},\ell) C_{\ell_{1}}^{(\mathrm{F})} C_{\ell_{2}}^{(\mathrm{EE})},
\label{FR2}
\end{equation}
where ${\mathcal G}(\ell_{1}, \ell_{2},\ell)$ is a function of the multipoles containing a Clebsh-Gordon coefficient 
 \cite{F2,gk2}, $C_{\ell}^{(\mathrm{EE})}$ is the angular power spectrum 
of the polarization autocorrelations and\footnote{For illustrative purposes we will limit our attention on the 
case $n_{\mathrm{B}} > 1$; in this situation $A_{\mathrm{B}}=  (2\pi)^{n_{\mathrm{B}}-1} \Gamma((n_{\mathrm{B}}-1)/2) 
B_{\mathrm{L}}^2$.}
\begin{equation}
C_{\ell}^{(\mathrm{F})} = 30.03 \,\,\overline{\Omega}_{\mathrm{BL}}
\biggl(\frac{\nu}{\nu_{\mathrm{max}}}\biggr)^{-4} \biggl(\frac{k_{0}}{k_{\mathrm{L}}}\biggr)^{n_{\mathrm{B}} -1} \frac{\ell (\ell + 1) (2\pi)^{n_{\mathrm{B}} -1}}{\Gamma\biggl(\frac{n_{\mathrm{B}} -1}{2}\biggr)}  \frac{\Gamma\biggl(\frac{5 - n_{\mathrm{B}}}{2}\biggr) \Gamma\biggl( \ell + \frac{n_{\mathrm{B}}}{2} - \frac{3}{2}\biggr)}{\Gamma\biggl(\frac{6 - n_{\mathrm{B}}}{2}\biggr) \Gamma\biggl(\frac{7}{2} + \ell - \frac{n_{\mathrm{B}}}{2}\biggr)},
\label{FR3}
\end{equation}
where $\overline{\Omega}_{\mathrm{BL}} = B_{\mathrm{L}}^2/(8\pi \overline{\rho}_{\gamma})$. 
In Fig. \ref{Figure2} the BB angular power spectra are reported. In the plot at the left the dashed line 
shows the result obtainable from Eq. (\ref{FR2}) in the case the $C_{\ell}^{(\mathrm{EE})}$ would 
be the one arising in the context of the $\Lambda$CDM adiabatic mode. 
The WMAP 5-year data (see, in particular, \cite{WMAP5}) imply that, when averaged over $\ell= 2-6$,
$\ell (\ell +1) C_{\ell}^{(\mathrm{BB})}/(2\pi) < 0.15 (\mu\mathrm{K})^2$ ( $95\,\%$ C.L.). 
The putative constraint  of \cite{WMAP5} does not make reference 
to a specific frequency. So it should be imposed at the lowest frequency channel. The lowest available 
frequency for this purpose would be for $27$ GHz. The preceding frequency (i.e. $23$ GHz) 
has been used as a foreground template and, consequently, the EE and BB
power spectra have not been freed from the foreground contamination.
We therefore choose to set the bound for a minimal frequency of $30$ GHz since this is 
not only intermediate between the KKa and KQ bands of the WMAP experiment 
but it is also the putative (lowest) frequency of the Planck experiment \cite{planck}.
In Fig. \ref{Figure3} (plot at the left) the full, dashed and dot dashed lines refer, respectively, to the cases 
of $\nu =30$ GHz, $\nu = 100$ GHz and $\nu= \nu_{\mathrm{max}}$. 
In the right plot of Fig. \ref{Figure3} the magnetized EE and BB power spectra are 
compared with the B-modes from the lensing of CMB anisotropies and from the B-modes induced 
by the tensor modes (in the case of tensor-to-scalar ratio $0.3$ which is the best fit value of the 
$\Lambda$CDM model plus tensors to the WMAP 5-year data). As it is 
apparent from Fig. \ref{Figure3} the constraints on the B-mode 
are not stringent for the magnetized background and are safely satisfied 
by a nG field at the epoch of the gravitational collapse \footnote{The WMAP collaboration 
reports also limits on axion-induced birefringence. As discussed in \cite{gk2} these bounds are
obtained by assuming that the birefringence is independent on the frequency of the incoming polarization (which is not true in the case of the magnetic field). Furthermore, the rate of axion-induced birefringence is fully homogeneous which is opposite 
to the case considered here.}. The constraints on the height of the acoustic peaks are comparatively more stringent \cite{gk,gk2}.

The obtained results suggest that multifrequency measurements of the CMB temperature and polarization within 
different channels will permit, for instance with Planck \cite{planck}, an accurate scrutiny of the possible 
presence of magnetized birefringence. For this purpose, the scaling properties of the temperature and polarization autocorrelations 
in different frequency channels should be analyzed and compared. 
While $C_{\ell}^{(\mathrm{BB})}$ should scale as $\nu^{-4}$,  the EE and TT power spectra will be frequency 
independent \cite{gk2}.

K.E.K. is supported by the ``Ram\'on y Cajal''  program and by the grants FPA2005-04823, FIS2006-05319 and CSD2007-00042 of the Spanish Science Ministry.

\end{document}